\documentclass[twocolumn,aps,prl,showpacs,preprintnumbers, superscriptaddress, amsmath,amssymb]{revtex4}

\usepackage{stmaryrd} 
\usepackage{mathtools} 
\usepackage{stmaryrd} 

\newcommand {\Tr}{{\rm Tr}\,}

\usepackage{amsmath}
\usepackage{amsfonts}

\usepackage{amssymb}
\usepackage{graphicx}

\begin{document}

\author{Andre G. Campos}
\email{agontijo@princeton.edu}

\author{Denys I. Bondar}
\email{dbondar@princeton.edu}

\author{Renan Cabrera}

\author{Herschel A. Rabitz}
\affiliation{Department of Chemistry, Princeton University, Princeton, NJ 08544, USA} 

\title{Making Distinct Dynamical Systems Appear Spectrally Identical}

\date{\today}

\begin{abstract}
We show that a laser pulse can always be found that induces a desired optical response from an arbitrary dynamical system. As illustrations,  driving fields are computed to induce the same optical response from a variety of distinct systems (open and closed, quantum and classical). As a result, the observed induced dipolar spectra without detailed information on the driving field is not sufficient to characterize atomic and molecular systems. The formulation may also be applied to design materials with specified optical characteristics.  These findings reveal unexplored flexibilities of nonlinear optics.
 \end{abstract}


\maketitle

\emph{Introduction}. One system imitating another different system, known as mimicry, abounds in the sciences. For example, in biology \cite{gust2001mimicking,cremer2002chemical,van2007expanding}, different  species 
often change their appearance in order to hide from predators. 
In material science \cite{leininger2000self,zhang2005self,mitov2002long,whitesell1994molecular} and chemistry \cite{ChirikScience327-794(2010),gust1993molecular,de1997signaling,della2014mimicry,rabong2014nxo},  
simpler and cheaper compounds one sought to mimic the properties of more complex and 
 expensive materials.
In this Letter, we introduce the method of \textit{Spectral Dynamic Mimicry} (SDM) bringing imitation into the domain of optics via quantum control. Thereby, SDM may be viewed as realizing an aspect of the alchemist dream to make different elements or materials look alike, albeit for the duration of a control laser pulse.

\begin{figure}
  \includegraphics[width=1\hsize]{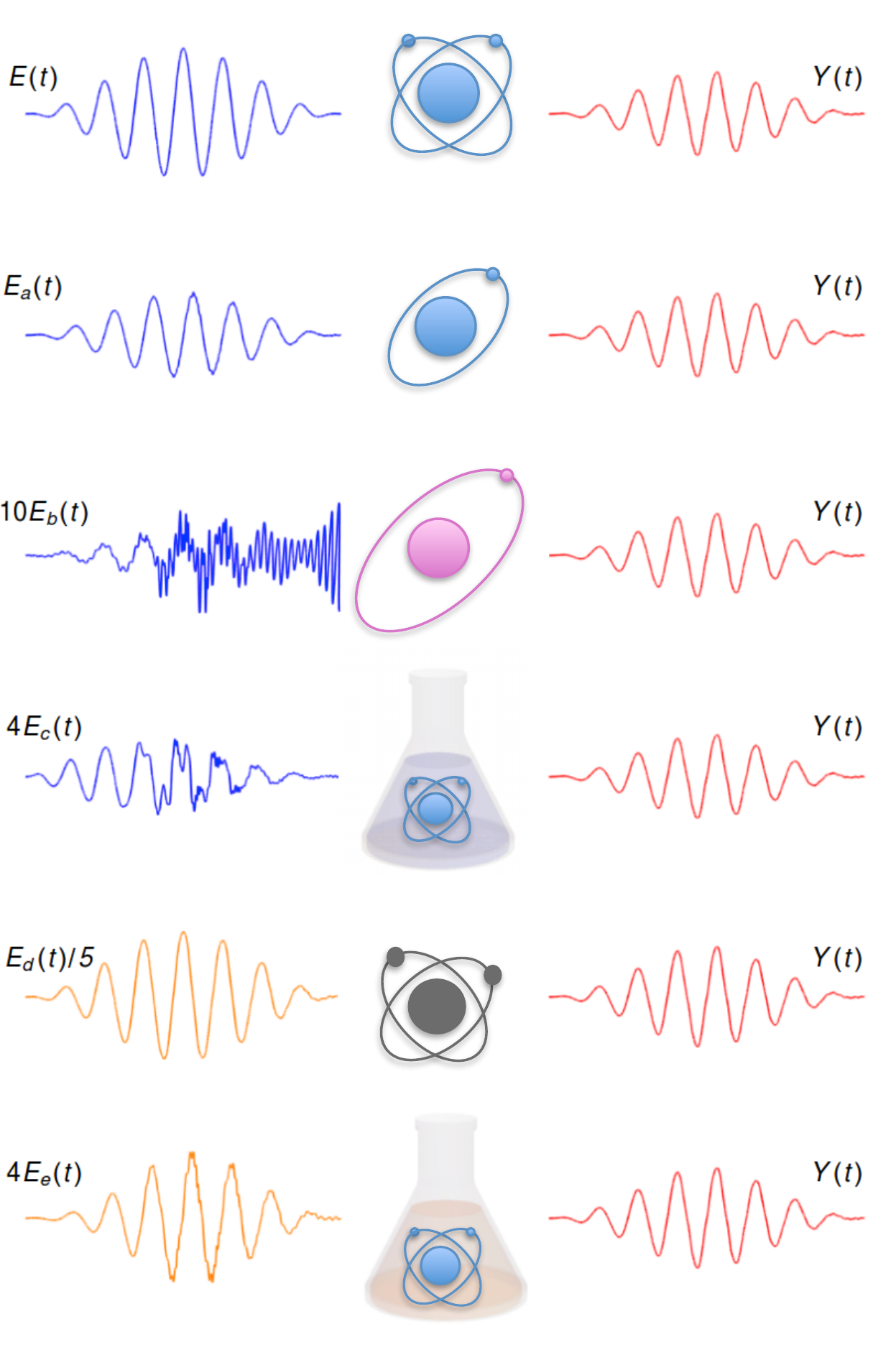}
  \caption{ The control fields $E_a(t)$, $E_b(t)$, $E_c(t)$, $E_d(t)$ and $E_e(t)$ induce the same nonlinear optical response $Y(t)$ on: closed quantum systems (a) and (b), (c) an open quantum system, (d) an closed classical system and (e) an open classical system. In (a) and (b), the system is a hydrogen atom initially prepared in the ground and first excited states, respectively. $E_b(t)$, $E_d(t)$ and $E_e(t)$ are scaled for comparison with the first field $E(t)$ that is applied to a model of an argon atom to produce the induced dipolar spectra $Y(t)$ ,which is used for tracking in the remaining cases. Compare with Fig. \ref{FigAllExamples} (f) for the importance of small differences in the control fields, which are not apparent in the time domain.} 
     \label{figDiagram}
\end{figure} 

\emph{Summary of results}.  We want the induced dipole spectra (IDS) of an $N$-electron system, $\vec{y}(t)=\sum_{k=1}^N\langle \vec{\hat{\boldsymbol{ x}}}_k \rangle$, to follow (i.e., track) a predefined time-dependent vector $\vec{Y}(t)$; atomic units (a.u.) with $\hbar=m=e=1$ are used throughout. In particular, assuming that $\vec{y}(t) = \vec{Y}(t)$ at some time moment $t$, then the control field $\vec{E}(t+dt)$ enforcing $\vec{y}(t+dt) = \vec{Y}(t+dt)$ at the next time step $t+dt$ is given by 
\begin{align}
	&\vec{E}(t + dt) = -\frac{4}{Ndt}\left[ \left\langle\sum_{k=1}^N \vec{\hat{\boldsymbol{ p}}}_k \right\rangle - \frac{\vec{Y}(t+dt) - \vec{Y}(t)}{dt} \right] \nonumber\\
	&\quad+\frac{2}{N}\left\langle \sum_{k=1}^{N}\mathbf{\nabla}_k V_k(\vec{\hat{ \boldsymbol{ x}}}_k)\right\rangle-\frac{2}{N}\left\langle \sum_{k=1}^N\vec{A}_k\right\rangle - \vec{E}(t) + O\left( dt \right),
	\label{EqField3DNp}
\end{align}
where $dt$ is an infinitesimal time increment, $\left\langle -\nabla_k V(\vec{\hat{\boldsymbol{ x}}}_k) \right\rangle(t)$ and  $\langle \vec{A}_k\rangle(t)$ describe the interaction with a potential force and an environment, respectively (see Sec. I of Supplemental Material \cite{SupplementalM} for details, which includes Ref. \cite{bandrauk1992higher}). The state (i.e., the density matrix and the probability distribution in the quantum and classical cases, respectively) determining the expectation values is propagated to the next time moment via the corresponding equation of motion (see, e.g., Table \ref{Table}) using $\vec{E}(t+dt)$. Having calculated $\vec{E}(t)$ for all times, the dynamical equation is used to verify satisfaction of the tracking condition $\vec{y}(t) = \vec{Y}(t)$. 
 
Since Eq. (\ref{EqField3DNp}) has exactly the same structure of the single particle case, we will study systems with single-electron excitation (i.e., $N=1$) in one spatial dimension.  In this case, Eq. (\ref{EqField3DNp}) takes the form
\begin{align}
 	E(t + dt) = &-\frac{4}{dt}\left[ \left\langle \hat{\boldsymbol{ p}} \right\rangle(t) - \frac{Y(t+dt) - Y(t)}{dt} \right] \nonumber\\
	&+ 2\left\langle V'(\hat{\boldsymbol{ x}}) \right\rangle(t)  - 2\langle A \rangle(t) - E(t) + O\left( dt \right),
	\label{Eq_Field}
\end{align}
where $\left\langle -V'(\hat{\boldsymbol{ x}}) \right\rangle$ and  $\langle A \rangle$ are specified in Table \ref{Table} for widely used models. The described scheme constitutes SDM, as the distinct physical systems in Fig. \ref{figDiagram} produced the same $Y(t)$, yet the resulting control fields calculated from Eq. (\ref{Eq_Field}) are unique once the system's initial state is supplied.

The physical meaning of Eq. (\ref{Eq_Field}) is that a desired polarizability can be induced from any dynamical system as long as no constraints are imposed on the driving laser field. In this fashion, the IDS of \textit{any} two atomic or molecular systems can be made identical by applying the specific required pulse shapes. 
\begin{center}
\begin{table*}
	\begin{tabular}{|c|c|c|p{4.8cm}|}
	\hline
	Type of system& $\langle O\rangle$  & $\langle A\rangle(t)$ & Equations of motion    \\ 
	\hline
	\hline
	Closed Quantum & $ \Tr[\hat{\rho} \hat{O}]$ & $0 $ & $i\hbar d\hat{\rho}/dt=[\hat{H},\hat{\rho}]$ \\
	\hline
	Open Quantum  & $\Tr[\hat{\rho} \hat{O}]$ & $-2\gamma\langle \hat{\boldsymbol{ p}}\rangle$ & $\mbox{Caldeira-Legget equation, Eq. (\ref{eqOpenCaldeira})}$  \\
	\hline
	Closed Classical &$\frac{1}{N}\sum_{i=1}^N O(x_i(t),p_i(t))$ &$0$ & $\mbox{Newton's equations, Eq. (\ref{NewtonEqs})}$\\
	\hline
	Open Classical &$\int dxdpO(x,p)\rho(x,p,t)$ &$- 2\gamma\langle p\rangle$ & $\mbox{Fokker-Planck equation, Eq. (\ref{FokkerPlanck})}$\\
	\hline
	\end{tabular}
\caption{The dynamical systems, averages and equations of motions analyzed in this Letter. In all quantum cases, we have $\hat{H} = \hat{H}_0-\hat{\boldsymbol{ x}}E(t)$, where $\hat{H}_0 = \hat{\boldsymbol{ p}}^2/2 + V(\hat{\boldsymbol{ x}})$. The atoms are represented as a single active electron moving in the field of a soft-Coulomb potential $V(x) = -Z_e/\sqrt{x^2+a^2}$, where $Z_e$ is the effective charge of the atom and both $Z_e$ and $a$ are chosen such that the experimental value of the ionization potential is reproduced. In the open classical case, $\rho(x,p,t)$ correspond to the particle's probability density in the phase space.}\label{Table}
\end{table*}
\end{center}

Such versatility of SDM is due to the fact that the induced dynamics takes advantage of the continuum. The IDS $y(t)$, as an expectation value of $\hat{\boldsymbol{x}}$, can attain arbitrary values only if the coordinate $\hat{\boldsymbol{x}}$ is unrestricted.  Moreover, if a strong field $E(t)$ is required to match an IDS, then $E(t)$ may induce ionization necessitating the coupling to the continuum. 
Mathematically, this means that $\hat{\boldsymbol{x}}$ and $\hat{\boldsymbol{p}}$ need to act in an \emph{infinite} dimensional Hilbert space. 

Equation (\ref{Eq_Field}) is a special case of 
tracking control \cite{gross1993inverse}:
Given a desired target $Z(t)$, find the control $E(t)$ such that $Z(t) = z(t)$ with $z(t)=\langle \hat{O}\rangle$ for a chosen observable $\hat{O}$. For simplicity, consider a closed quantum system ($\hat{A}=0$) with the hamiltonian $\hat{H}(t) = \hat{H}_0 - E(t) \hat{\mu}$ and $[\hat{O},\hat{\mu}]=0$. The corresponding Ehrenfest theorem then reads
\begin{align}\label{EqTrackingControl}
	\frac{d^2Z(t)}{dt^2} =-\left\langle[\hat{H}_0,[\hat{H}_0,\hat{O}]] \right\rangle+\left\langle[\hat{ \mu},[\hat{H}_0,\hat{O}]] \right\rangle E(t).
\end{align}
Given $Z(t)$, Eq. (\ref{EqTrackingControl}) is solved with respect to the unknown $E(t)$. Tracking control has been typically applied to finite-level quantum systems. In this case, $\langle[\hat{ \mu},[\hat{H}_0,\hat{O}]] \rangle$ may vanish at some time $t$, leading to a singularity in $E(t)$. There is no general way to prevent these singularities in finite dimensional tracking irrespective of the form of the \emph{finite} dimensional matrices $\hat{H}_0$, $\hat{\mu}$, and $\hat{O}$
\footnote{
A general condition to avoid the singularity of the control field $E(t)$ in Eq. (\ref{EqTrackingControl}) is 
either the positive- or negative-definiteness of the operator $[\hat{ \mu},[\hat{H}_0,\hat{O}]]$. However, this is impossible in a finite dimensional setting since no finite matrix of the form $[\hat{A}, \hat{B}]$ is sign definite. The latter follows from the definition of the trace as the sum of eigenvalues, and the general observation that $\Tr [\hat{A}, \hat{B}] = 0$ implying that either $\hat{A}$ and $\hat{B}$ commute or $[\hat{A}, \hat{B}]$ has both positive and negative eigenvalues.
}. 
Note that SDM is free of such singularities by construction
\footnote{
Assume that $\hat{O} = O(\hat{\boldsymbol{x}})$, $\hat{\mu} = \mu(\hat{\boldsymbol{x}})$, and $\hat{H}_0 = \hat{\boldsymbol{p}}^2 / 2$, where $\hat{\boldsymbol{x}}$ and $\hat{\boldsymbol{p}}$ are self-adjoint operators acting on the infinite dimensional Hilbert space. Since $[\hat{ \mu},[\hat{H}_0,\hat{O}]]=d\hat{\mu}/dx \, d\hat{O}/dx$, the observable $\hat{O}$ can be tracked with no singularities in the control field E(t) [Eq. (\ref{EqTrackingControl})] if the function $d\mu(x)/dx \, dO(x)/dx$ is sign definite. Note that SDM [Eq. (\ref{Eq_Field})] corresponds to the simplest case with $\hat{O} = \hat{\mu} = \hat{\boldsymbol{x}}$.
}. 

\begin{figure}
  \includegraphics[width=1\hsize]{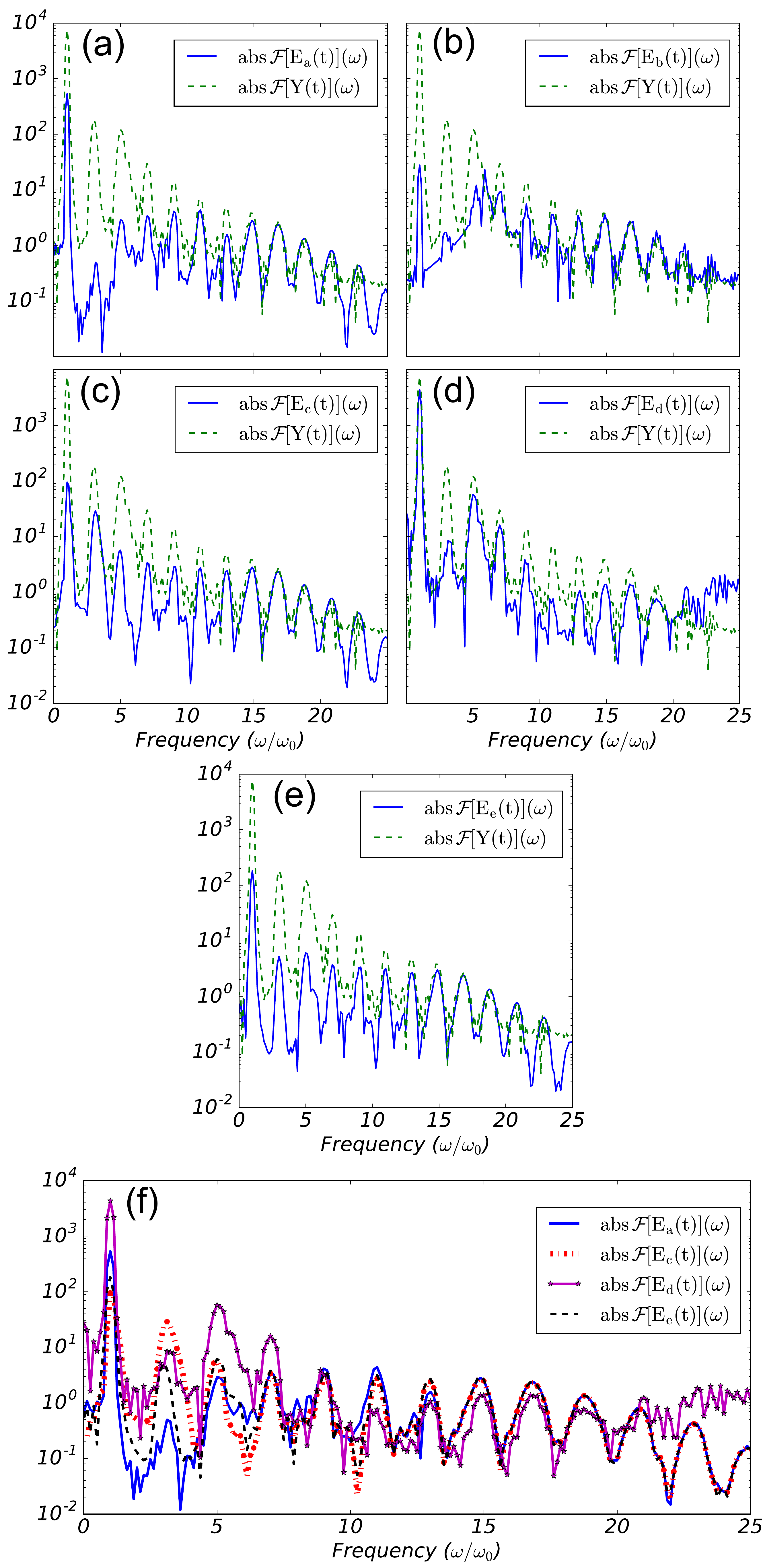}
  \caption{
 (color online) The control fields $E_a(t)$, $E_b(t)$, $E_c(t)$, $E_d(t)$, $E_e(t)$ and the target IDS $Y(t)$ from Fig. \ref{figDiagram} in frequency domain. $\rm abs\, \mathcal{F}[\cdot]$ denotes the absolute value of the Fourier transform. The $\omega_0$ is the carrier frequency of $E(t)$ in Fig. \ref{figDiagram}.
  } 
  \label{FigAllExamples}
\end{figure}

According to Eq. (\ref{Eq_Field}), the control field is shaped in time domain, thus possibly introducing high frequency components beyond the target response bandwidth $\omega/\omega_0\gtrsim23$, as seen in Fig. \ref{FigAllExamples}. However, those high frequencies are not important for the dynamics, since removing all frequencies outside the target response  bandwidth, (i.e., for $\omega/\omega_0\gtrsim23$) in the tracking fields $E_a(t)$, $E_c(t)$ and $E_d(t)$, does not significantly affect the tracking condition $y(t) = Y(t)$. Moreover, SDM is robust to the presence of multiplicative noise in the tracking (see Sec. II of Supplemental Material \cite{SupplementalM} for details, which includes Ref. \cite{bandrauk1992higher}).

Equation (\ref{Eq_Field}) describes a broad variety of dynamical systems (Table \ref{Table}). As illustrations, we apply SDM to the following models: (a) and (b) closed quantum systems governed by the von Neumann equation, (c) open quantum systems modeled by the Caldeira-Legget master equation \cite{caldeira1983path,caldeira1983quantum},  (d) classical closed systems obeying Newton's equations, and (e) open classical systems described by the Fokker-Planck equation \cite{gardiner1985handbook}. 
In all these cases, we track the target $Y(t)$, which is obtained as an IDS of an isolated argon atom treated as having one electron responding to a band limited field 
$E(t)=0.04\cos(\omega_0 t)f(t)$ of central frequency $\omega_0=0.06$ (a.u) (756 nm) and  the envelope $f(t)=\cos^2 (\pi t/(2t_f))$, where the final propagation time is $t_f=8\pi/\omega_0$. Figure \ref{FigAllExamples} depicts the spectrum of $Y(t)$ exhibiting high harmonic generation (HHG) \cite{PhysRevLett.71.1994,RevModPhys.80.117}.

We employ the single active electron approach \cite{PhysRevA.56.3028} to model atomic systems throughout. Hence, an atom is represented by a single particle moving in the field of a soft-Coulomb potential 
\begin{align}\label{SoftCoulombPot}
	V(x) = \frac{-Z_e}{\sqrt{x^2+a^2}},
\end{align}
where the effective charge $Z_e \approx 1$ and radius $a$ is adjusted such that the ground state energy in each case matches the experimental ionization potential. For example, $Z_e=1$ and $a^2=2$ (a.u.) models a hydrogen atom, while  $Z_e=1$ and $a^2=1.37$ (a.u.) are used for argon. Since there is no degeneracy in the spectrum of one dimensional quantum systems, the eigenstates are labelled only by the principal quantum number $n$. 

\emph{Making two closed quantum systems look alike}. As a first example of SDM, we make hydrogen `look like' argon by matching their IDS. We find the shape of the laser field $E_a(t)$ [Eq. (\ref{Eq_Field})] that induces an optical response $Y(t)$ in a hydrogen atom initially in the ground state ($n=1$), modeled as a closed quantum system. Superimposing the Fourier transforms of $E_a(t)$ and $Y(t)$ in Fig. \ref{FigAllExamples} (a), we note that  the tracking field bandwidth is broader than the target bandwidth. This trend is observed in all examples presented in this Letter. Moreover, the third harmonic ($3\omega_0$) in the tracking field is an order of magnitude smaller than the same frequency in $Y(t)$. A further analysis reveals that the induced response is at best weakly dependent of the third harmonic in the driving field. Thus, the $3\omega_0$ generation in hydrogen occurs via parametric down conversion \cite{Boyd}.

As mentioned earlier, the control fields calculated using SDM are unique once the system's initial state is specified. For the control field $E_a(t)$ shown in Fig. \ref{figDiagram}, the hydrogen atom was initially in the ground state; however, a very different control field is required ($E_b(t)$ in Figs. \ref{figDiagram} and \ref{FigAllExamples} (b)) if the hydrogen atom is initially in the first excited state ($n=2$).  The amplitude of $E_b(t)$ is nearly a factor of ten smaller than $E_a(t)$ (see Fig. \ref{figDiagram}) since the energy gap between the ground and first excited states in the hydrogen atom is approximately half the ionization potential. Moreover, for the hydrogen atom in the first excited state, single photon ionization takes place for $\omega/\omega_0\approx5$, whereas parametric down conversion dominates the dynamics for $\omega/\omega_0\lesssim7$ (see Fig. \ref{FigAllExamples} (b)). For higher frequencies where $\omega/\omega_0\gtrsim7$, linear response takes place.

\emph{Making open and closed quantum systems look alike}. The effects of energy damping and dephasing
are commonly modeled by the Caldeira-Legget master equation \cite{caldeira1983path,caldeira1983quantum}
\begin{align}
	i\frac{d\hat{\rho}}{dt}=[\hat{H},\hat{\rho}] + i\mathcal{L}[\hat{\rho}],	\quad
	\mathcal{L}[\hat{\rho}]=-i\gamma[\hat{\boldsymbol{ x}},[\hat{\boldsymbol{ p}},\hat{\rho}]]-\chi[\hat{\boldsymbol{ x}},[\hat{\boldsymbol{ x}},\hat{\rho}]],
\label{eqOpenCaldeira}
\end{align}
where $\chi=2\gamma kT$. Using Eq. (\ref{Eq_Field}) with the damping term $\langle A \rangle = -2\gamma\langle\hat{\boldsymbol{ p}}\rangle$ as specified in Table \ref{Table}, we find the control field $E_c(t)$ (Fig. \ref{FigAllExamples} (c)) that induces the optical response of the atomic argon, interacting with a dissipative environment, to match  the nonlinear spectra of the isolated argon $Y(t)$ shown in Fig. \ref{figDiagram}. The amplitude damping time and temperature were chosen as $\gamma^{-1}\approx$ 242 fs  and $T\approx$ 100 K, correspondingly.

According to dynamical decoupling  \cite{PhysRevA.58.2733,PhysRevA.71.022302,ban1998photon,cappellaro2006principles},
appropriately designed laser pulses can compensate for the interaction of a quantum system with the environment. Dynamical decoupling usually relies on a perturbative treatment of the environment; whereas, SDM [Eq. (\ref{Eq_Field})] is explicitly nonperturbative in both the field and environmental interactions. 

HHG is an important source for creating attosecond pulses. A weak HHG signal is often obtained by irradiating a low pressure inert gas with a band limited pulse \cite{RevModPhys.80.117}. The intensity of HHG is proportional to the gas concentration. However, the denser the gas, the less isolated the atomic system becomes, giving rise to decoherent dynamics, thus suppressing HHG \cite{Constant1999PhysRevLett.82.1668}. The presented SDM illustration shows that the HHG spectra of an isolated system can be reproduced even from an open system by properly pulse shaping the incident laser field.

\emph{Making closed classical and quantum systems look alike}. The position and momentum of an ensemble of $N$ classical particles obey Newton's equations 
\begin{align}
	\frac{d}{dt}x_i(t)=p_i(t),\quad
	\frac{d}{dt}p_i(t)=-V'(x_i(t)) + E(t),
	\label{NewtonEqs}
\end{align}
    where $V(x)$ is given by Eq. (\ref{SoftCoulombPot}). The ensemble's initial momentum and positions are randomly generated by the normal distribution with zero mean and standard deviation of $1$. According to Table \ref{Table}, the IDS of the classical system $y_d(t)$ is given by
$
	y_d(t)=\frac{1}{N}\sum_{i=1}^N\,x_i(t).
$
From Eq. (\ref{Eq_Field}), we find the control field $E_d(t)$ (Fig. \ref{FigAllExamples} (d)) that forces the IDS of the classical argon model [Eq. (\ref{NewtonEqs})] to match the IDS of the isolated argon $Y(t)$.

The dynamics underlying the classical IDS $y(t)$ induces nonlinear optical processes. In particular, the further the trajectory goes from the center of force (i.e., the origin), the more harmonics it yields. This can be seen from the following taylor expansion of Eq. (\ref{NewtonEqs})
\begin{align}
	\frac{d}{dt}p_i(t)\approx  E(t) - \frac{Z_e}{a^3}x_i(t)+\frac{3Z_e}{2a^5}x_i(t)^3 + O\left(x_i(t)^5\right).
	\label{NewtonEqsTaylor}
\end{align}
The first two terms on the right hand side correspond to a driven harmonic oscillator. Therefore, the trajectories closest to the origin only linearly respond to the control field $E(t)$; whereas, the trajectories farther away give rise to high harmonics. As can be seen in Fig. \ref{FigAllExamples} (f), the spectrum of the classical control field $E_d(t)$ deviates significantly from the previously obtained control fields $E_a(t)$, $E_c(t)$ and $E_e(t)$ for the quantum cases. As in system (a) of Figs. \ref{figDiagram} and \ref{FigAllExamples} (a), suppressing the third harmonic ($3\omega_0$) in the classical control field does not significantly affect the response. It is noteworthy that significantly nonlinear classical dynamics can be indistinguishable from quantum evolution \cite{Franco2010143,schwieters1993optimal,blackburn2016survey}.

\emph{Making open classical and closed quantum systems look alike}. The state of an open classical system can be specified by a positive probability distribution function $\rho=\rho(x,p,t)$ defined on a classical phase space. The dynamics of such a system is commonly modeled by the Fokker-Planck equation \cite{gardiner1985handbook}
\begin{align}
	\frac{\partial \rho}{\partial t} 
	= \left[-\frac{\partial}{\partial x}p+(V'(x)-E(t))\frac{\partial}{\partial p} +2\gamma\frac{\partial}{\partial p}p+D\frac{\partial^2}{\partial p^2}\right]\rho,
\label{FokkerPlanck}
\end{align}
where $D=0.01$ (a.u.) denotes a diffusion coefficient and $\gamma=0.001$ (a.u.) quantifies energy damping.

Following Ref. \cite{PhysRevA.92.042122}, we use Eq. (\ref{FokkerPlanck}) as a classical counterpart of the Caldeira-Legget Eq. (\ref{eqOpenCaldeira}), modeling the atomic argon interacting with a dissipative bath. From Eq. (\ref{Eq_Field}) with $\langle A \rangle=-2\gamma\langle p\rangle$ as specified in Table \ref{Table}, we find the control field $E_e(t)$ (Figs. \ref{figDiagram} and \ref{FigAllExamples} (e)) that forces the IDS of the argon classical model [Eq. (\ref{FokkerPlanck})] to match the IDS of the isolated argon $Y(t)$.

It is important to note the remarkable similarity between $E(t)$ and $E_e(t)$ in Fig. \ref{figDiagram}. In fact, for our particular example of open classical dynamics, the intensity of the IDS is proportional to the intensity of the control field $E_e(t)$. Furthermore, reducing the intensity of any individual frequency in the control field linearly decreases the intensity of the corresponding harmonic in the IDS without influencing the other frequencies. This shows that there is only a linear optical process taking place. Moreover, there are no cooperative effects between different frequency components -- a consequence of strong decoherence in the particular example of open classical dynamics considered here (see also Ref. \cite{Franco2010143}).

The spectrum of the tracking fields $E_a(t)$, $E_c(t)$, $E_d(t)$ and $E_e(t)$ are shown in Fig. \ref{FigAllExamples} (f). Subsequent analysis indicates that the optical responses for the closed (a) and open (c) quantum systems are nonlinear in the frequency ranges of $\omega/\omega_0\lesssim 13$ and $\omega/\omega_0\lesssim 7$, respectively. Similar to our simulations of the open classical system (e), laser-matter interactions described within the classical and quantum electrodynamics coincide in the linear response regime \cite{andrews2004virtual, PhysRevLett.115.110402Brumer}. In contrast, the closed classical system (d) displays strong nonlinear effects overall, as can be seen in Fig. \ref{FigAllExamples} (d) and (f). 

\emph{Conclusions}. We put forward the paradigm of SDM
in which laser fields are shaped to make any distinct dynamical system look identical spectrally to any other system. As a result, the observed IDS without any information on the driving field cannot be used to unambiguously characterize atomic and molecular systems.  

SDM can be applied to many important problems. For example, it can be seen as the opposite of \textit{Optimal Dynamic Discrimination} (ODD) which shows that nearly identical quantum systems may be distinguished by means of their dynamics induced by properly shaped laser pulses \cite{li2002optimal,goun2016photonic}. ODD has been experimentally confirmed for a number of nominally similar systems \cite{brixner2001photoselective,tkaczyk2008control,roth2009quantum,petersen2010shaped,rondi2011discriminating,rondi2012coherent}.  In future works, we plan to reformulate ODD as a tracking control problem (in the spirt of SDM) in order to propose novel methods for the concentration characterization of a mixture of complex molecular species with similar linear spectra. This problem is inspired by the challenges in the life sciences \cite{Constant1999PhysRevLett.82.1668,feng2000imaging,bagwell1993fluorescence,speicher1996karyotyping,perfetto2004seventeen}. ODD may also be used to find control fields that optimally discriminate between classical and quantum models of the same physical system, thereby shading light on ongoing discussions \cite{schwieters1993optimal,blackburn2016survey,Franco2010143}.
Moreover, being non-perturbative in both the control field and environment interactions, SDM offers a potential alternative to dynamical decoupling \cite{PhysRevA.58.2733,PhysRevA.71.022302,ban1998photon,cappellaro2006principles}. Furthermore, in the framework of SDM, HHG spectra of an isolated system can be induced from an open system by pulse shaping the incident laser field, providing an efficient way to generate bright HHG from dense atomic gases. In addition, the high degree of robustness to noise of the tracking fields (see Sec. II of Supplemental Material \cite{SupplementalM}, which includes Ref. \cite{bandrauk1992higher}) makes SDM suited for experimental applications. 

As a final remark, a recently experiment \cite{sommer2016attosecond} demonstrated the feasibility of simultaneous characterization of the control field as well as IDS, opening a possibility of SDM experimental realization.

\emph{Acknowledgments}. A.G.C. acknowledges financial support from NSF CHE 1464569, D.I.B. from DOE DE-FG02-02-ER-15344, R.C. from ARO W911NF-16-1-0014 and H.R. from DARPA W911NF-16-1-0062. A.G.C. was also supported by the Fulbright foundation. D.I.B. is also supported by 2016 AFOSR Young Investigator Research Program. We would like to acknowledge an anonymous referee for drawing our attention to Ref. \cite{blackburn2016survey}. 

A.G.C. and D.I.B. contributed equally to this work.

\begin{widetext}

\begin{center}
{\Large\bf Supplemental Material for: ``Making Distinct Dynamical Systems Appear Spectrally Identical''} 
\end{center}

\tableofcontents
\section{Derivation of Equation (1) from the main text}

Consider open system dynamics, with Hamiltonian for $N$ interacting particles having the form
\begin{align}
	\hat{H}(t) &= \sum_{k=1}^N\frac{\hat{\boldsymbol{p}}^2_k}{2} + \sum_{k=1}^{N}V_i(\vec{\hat{\boldsymbol x}}_k) - \sum_{k=1}^{N}\vec{\hat{\boldsymbol x}}_k\cdot \vec{E}(t)+\frac{1}{2}\sum_{k,j=1}^{N}U(|\vec{\hat{\boldsymbol x}}_k-\vec{\hat{\boldsymbol x}}_j|), 
\end{align}
where $V_i(\vec{\hat{\boldsymbol x}}_k)$ corresponds to the interaction between each particle of the system and an external potential, $U(|\vec{\hat{\boldsymbol x}}_k-\vec{\hat{\boldsymbol x}}_j|)$ describes the inter-particle interaction and $\vec{E}(t)$ is the applied laser field. The dynamical equation for the open system dynamics with the above Hamiltonian is
\begin{align}
	\frac{d}{dt} \hat{ \rho}(t) = \mathcal{L}[\hat{\rho}(t)],
\end{align}
with
\begin{align}
	\mathcal{L}[\odot] = -i[\hat{H}(t), \odot ]  + \mathcal{D}[\odot],
\end{align}
where $\mathcal{D}[\hat{\rho}(t)]$ describes the coupling between the system and the environment. The time evolution of the density matrix from the instant $t$ to the next instant $t+dt$ in the super-operator formalism is
\begin{align}
	\hat{\rho}(t+dt) = \mathcal{T} e^{\int_{t}^{t+dt} \mathcal{L}[\odot]d\tau }\hat{\rho}(t),
\end{align}
where $dt$ is an infinitesimal time increment and $ \mathcal{T}$ is the time-ordering operator.

Given $\vec{Y}(t) = \sum_{k=1}^{N}\left\langle \vec{\hat{\boldsymbol x}}_k \right\rangle(t)$ at time $t$ , we would like to find the control field $E(t+dt)$ such that
\begin{align}
	\vec{Y}(t+dt) &= \sum_{k=1}^{N}\left\langle \vec{\hat{\boldsymbol x}}_k\right\rangle(t+dt) \notag\\
	&=  \sum_{k=1}^{N}{\rm Tr} \left[ \vec{\hat{\boldsymbol x}}_k \hat{\rho}(t + dt) \right]
	=  \sum_{k=1}^{N}{\rm Tr} \left[\vec{\hat{\boldsymbol x}}_k\,  \mathcal{T} e^{\int_{t}^{t+dt} \mathcal{L}[\odot] d\tau }\hat{\rho}(t) \right] \notag\\
	&=  \sum_{k=1}^{N}{\rm Tr} \left[ \vec{\hat{\boldsymbol x}} _ke^{\int_{t}^{t+dt} \mathcal{L}[\odot]d\tau }\hat{\rho}(t)\right]+ O\left( dt^3 \right) \notag\\
	&= \sum_{k=1}^{N} {\rm Tr} \left[ \vec{\hat{\boldsymbol x}}_k e^{dt\mathcal{L}_D[\odot]} \hat{\rho}(t) \right]+ O\left( dt^3 \right) \notag\\
	&=\sum_{k=1}^{N}\left\langle \vec{\hat{\boldsymbol x}}_k \right\rangle(t) + \sum_{k=1}^{N}\ {\rm Tr} \left[\vec{\hat{\boldsymbol x}}_k\left( dt\mathcal{L}_D[\hat{\rho}(t)] +\frac{dt^2}{2}\mathcal{L}_D[\mathcal{L}_D[\hat{\rho}(t)]]\right) \right] + O\left( dt^3 \right) \notag\\
	&= \sum_{k=1}^{N}\left\langle \vec{\hat{\boldsymbol x}}_k \right\rangle(t) + \sum_{k=1}^{N}\ {\rm Tr} \left[\left( dt\mathcal{L}_D^\dagger[\vec{\hat{\boldsymbol x}}_k] +\frac{dt^2}{2}\mathcal{L}_D^\dagger[\mathcal{L}_D^\dagger[\vec{\hat{\boldsymbol x}}_k]]\right) \hat{\rho}(t)\right] + O\left( dt^3 \right) \notag\\
\end{align}
where
\begin{align}
	\mathcal{L}_D[\odot]  = -\frac{i}{2}[\hat{H}(t+dt)+\hat{H}(t),  \odot ]  + \mathcal{D}[\odot] ,
\end{align}
and the factor of $2$ comes from using the midpoint rule to approximate the integral over time (see, e.g., Ref. \cite{bandrauk1992higher}). We also used the following definition of the adjoint 
\begin{align}
	{\rm Tr}\left[\hat{\vec{\boldsymbol x}}_k\mathcal{L}_D\left[\hat{\rho}(t)\right]\right] = {\rm Tr}\left[\mathcal{L}_D^\dagger\left[\hat{\vec{\boldsymbol x}}_k\right]\hat{\rho}(t)\right],
\end{align}
where 
\begin{align}
	\mathcal{L}_D^\dagger[\vec{\hat{\boldsymbol x}}_k] = \frac{i}{2}[\hat{H}(t+dt)+\hat{H}(t),  \vec{\hat{\boldsymbol x}}_k ]  + \mathcal{D}^\dagger[\vec{\hat{\boldsymbol x}}_k].
\end{align}
Let us assume that $\mathcal{D}^\dagger[\vec{\hat{\boldsymbol x}}_k]=0$, since all examples studied in the main text have this property. In this case, we have
\begin{align}
 	\mathcal{L}_D^\dagger[\mathcal{L}_D^\dagger[\vec{\hat{\boldsymbol x}}_k]] = \frac{i}{2}[\hat{H}(t+dt)+\hat{H}(t),\vec{\hat{\boldsymbol p}}_k] + \mathcal{D}^\dagger[\vec{\hat{\boldsymbol p}}_k]=-\vec{\nabla}_kV_k(\vec{\hat{\boldsymbol x}}_k) + \frac{1}{2}(\vec{E}(t+dt)+\vec{E}(t)) + \mathcal{D}^\dagger[\vec{\hat{\boldsymbol p}}_k],
\end{align}
since $-\sum_{k,j=1}^{N}\vec{\nabla}_kU(|\vec{\hat{\boldsymbol x}}_k-\vec{\hat{\boldsymbol x}}_j|)=\sum_{k,j=1}^{N}\vec{F}_{kj}=0$ due to Newton's third law, $\vec{F}_{jk}=-\vec{F}_{kj}$.

Thus, finally
\begin{align}
	\vec{E}(t + dt) = -\frac{4}{Ndt}\left[ \sum_{i=1}^{N}\left\langle \vec{\hat{\boldsymbol p}}_i \right\rangle(t) - \frac{\vec{Y}(t+dt) - \vec{Y}(t)}{dt} \right] 
	+\frac{2}{N} \sum_{i=1}^{N}\left\langle\vec{\nabla}_iV_i(\vec{\hat{\boldsymbol x}}_i) \right\rangle(t) -\frac{2}{N}\sum_{k=1}^N\langle\vec{\hat{A}}_k\rangle - \vec{E}(t) + O\left( dt \right),
	\label{EqEf1}
\end{align}
where $\vec{\hat{A}}_k =  \mathcal{D}^\dagger\left[\vec{\hat{\boldsymbol p}}_k\right]$. Performing a Taylor expansion in $\vec{Y}(t+dt)$, we get an equivalent result
\begin{align}
		\vec{E}(t + dt) = -\frac{4}{Ndt}\left[ \sum_{i=1}^{N}\left\langle \vec{\hat{\boldsymbol p}}_i \right\rangle(t) - \vec{Y'}(t) \right] 
	+\frac{2}{N}\vec{Y''}(t) + \frac{2}{N} \sum_{i=1}^{N}\left\langle\vec{\nabla}_iV_i(\vec{\hat{\boldsymbol x}}_i) \right\rangle(t) -\frac{2}{N}\sum_{k=1}^N\langle\vec{\hat{A}}_k\rangle - \vec{E}(t) + O\left( dt \right),
	\label{EqEf2}
\end{align}
which makes the connection with the Ehrenfest theorems explicit.

\section{Robustness to noise}
There are many different types of noise in lasers including phase, intensity and
frequency noise. In order to assess the
robustness of SDM, we consider the laser field $E_{a,b,c}(t)$ to be contaminated by 
a multiplicative noise. The model for the noisy laser field is
\begin{align}
	\bar{E}_{a,b,c}(t) = E_{a,b,c}(t)\left(1+W[0,\sigma_{a,b,c},t]\right),
	\label{noise}
\end{align}
where $W[\mu,\sigma,t]$ is a normally distributed  random variable with the mean $\mu=0$
and standard deviation $\sigma_{a,b}=0.02$ and $\sigma_{c}=0.05$, subscripts (a) and (b) denote closed and (c) open quantum systems ( see Fig. 1 in the main text). The signal to noise ratio
of the field's amplitude is
\begin{align}
SNR_{a,b,c} = \frac{\langle \bar{E}_{a,b,c}(t) \rangle}{\sqrt{Var[\bar{E}_{a,b,c}(t)]}}=\frac{1}{\sigma_{a,b,c}}.
\end{align}

The relative distance between the noise-free target IDS $Y(t)$ and the system's IDS $y(t)$ induced by the contaminated control field [Eq. (\ref{noise}] is estimated as
\begin{align}
	d(y(t),Y(t))^2 =\frac{ \int dt|y(t)-Y(t)|^2}{\int dt |Y(t)|^2}.
	\label{Euclidean}
\end{align}
Quantity (\ref{Euclidean}) evaluated for different noise realizations is shown in
Figs. \ref{EuclideanClosed} and \ref{EuclideanOpen} for the closed  and open quantum
system, respectively. In the open quantum system case,
we observe that the values of the relative distance (\ref{Euclidean}) calculated with noise fluctuate around the noise-free value. On the other hand, in Fig. \ref{EuclideanClosed} we note that the average 
value of the relative distance (\ref{Euclidean})  generated by the noisy control fields (blue solid
line) is approximately $18\%$ larger than the corresponding values for the noise-free tracking field (red dashed line).
\begin{figure}
  \includegraphics[width=1\hsize]{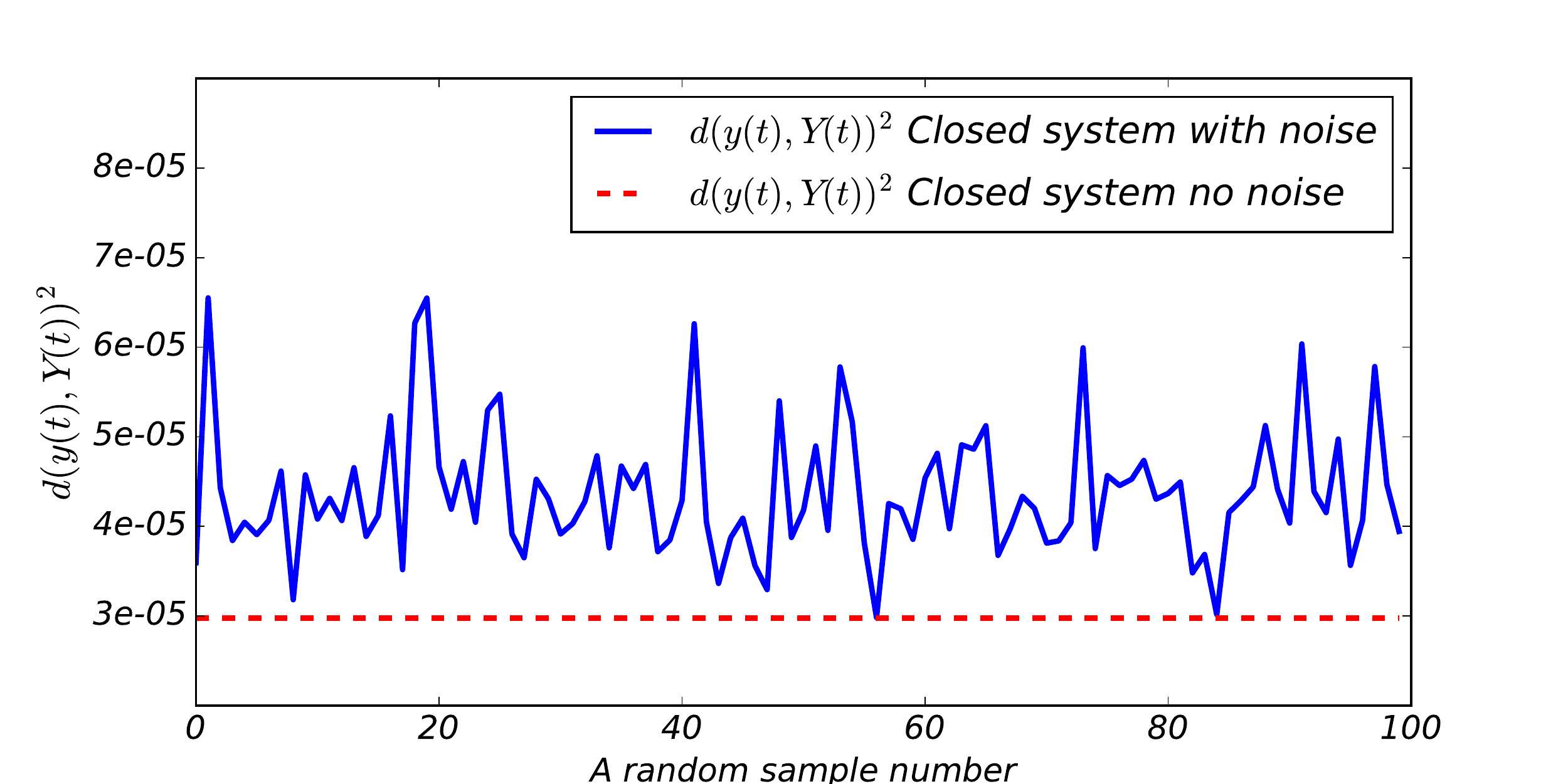}
  \caption{
 (color online) Plot of $d(y(t),Y(t))^2$ for the response generated by the closed system's control field with noise (solid blue line) and without noise (red dashed line). 
  } 
  \label{EuclideanClosed}
\end{figure} 
\begin{figure}
  \includegraphics[width=1\hsize]{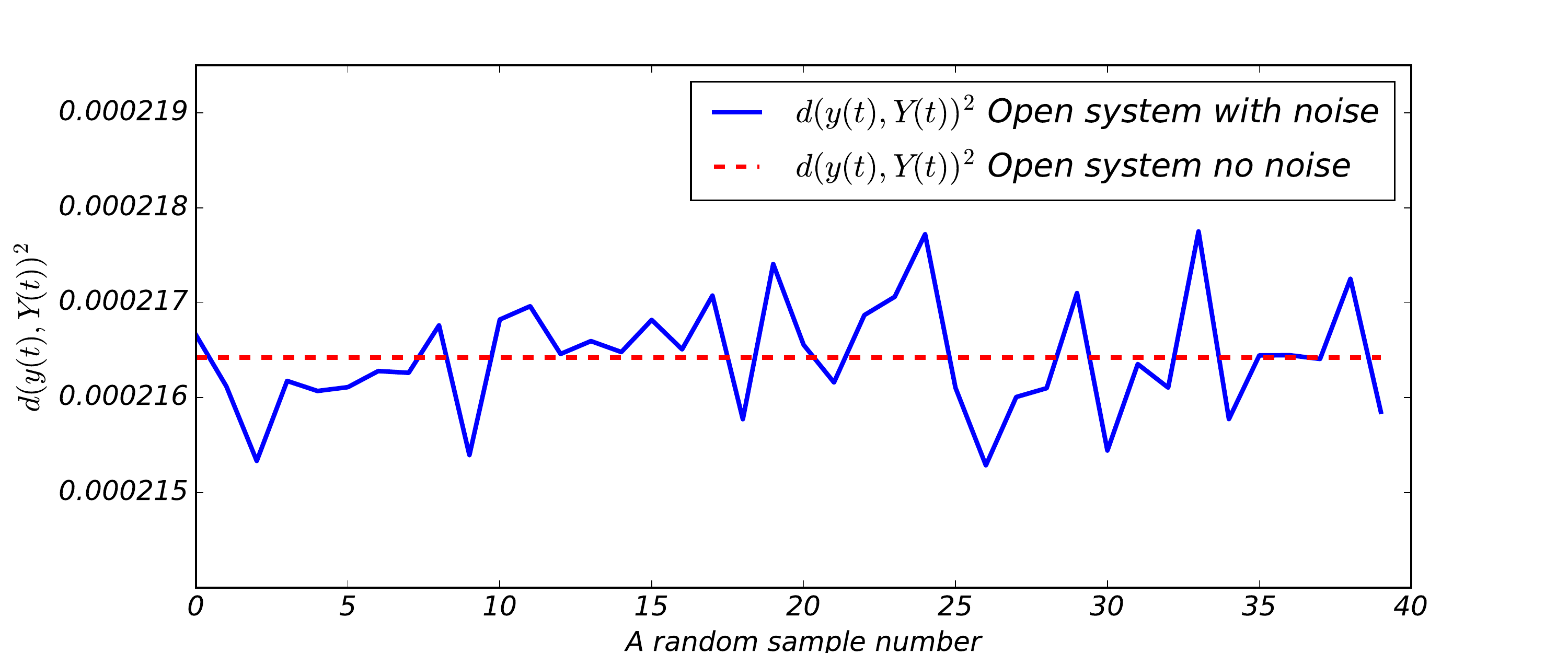}
  \caption{
 (color online) Plot of $d(y(t),Y(t))^2$ for the response generated by the open system's control field with noise (solid blue line) and without noise (red dashed line). } 
  \label{EuclideanOpen}
\end{figure} 
\begin{figure}
  \includegraphics[width=1\hsize]{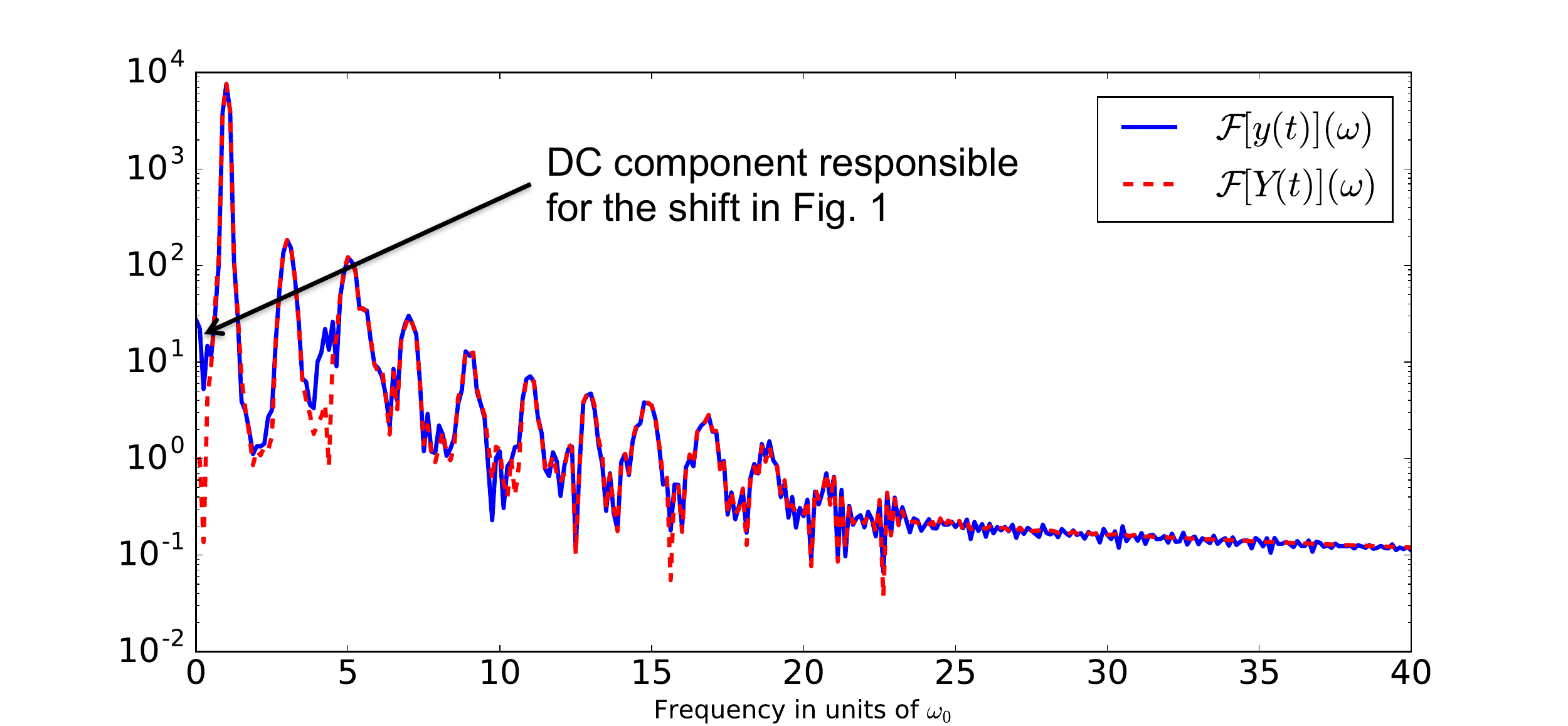}
  \caption{
 (color online) Plot of $\mathcal{F}[y(t)](\omega)$ for the closed system's IDS generated by the noisy field (blue, full curve) and the target IDS $\mathcal{F}[Y(t)](\omega)$ (red, dotted line). For this particular example, $d(y(t),Y(t))^2=6.5\times10^{-5}$. } 
  \label{responsenoise_freqDomain}
\end{figure} 
For the closed quantum systems, the noisy control field
generates a DC component in the IDS (see Fig. \ref{responsenoise_freqDomain} bellow), shifting the
average relative distance. 
\end{widetext}
\bibliography{bib-relativity}

\begin{thebibliography}{46}
\expandafter\ifx\csname natexlab\endcsname\relax\def\natexlab#1{#1}\fi
\expandafter\ifx\csname bibnamefont\endcsname\relax
  \def\bibnamefont#1{#1}\fi
\expandafter\ifx\csname bibfnamefont\endcsname\relax
  \def\bibfnamefont#1{#1}\fi
\expandafter\ifx\csname citenamefont\endcsname\relax
  \def\citenamefont#1{#1}\fi
\expandafter\ifx\csname url\endcsname\relax
  \def\url#1{\texttt{#1}}\fi
\expandafter\ifx\csname urlprefix\endcsname\relax\def\urlprefix{URL }\fi
\providecommand{\bibinfo}[2]{#2}
\providecommand{\eprint}[2][]{\url{#2}}

\bibitem[{\citenamefont{Gust et~al.}(2001)\citenamefont{Gust, Moore, and
  Moore}}]{gust2001mimicking}
\bibinfo{author}{\bibfnamefont{D.}~\bibnamefont{Gust}},
  \bibinfo{author}{\bibfnamefont{T.~A.} \bibnamefont{Moore}}, \bibnamefont{and}
  \bibinfo{author}{\bibfnamefont{A.~L.} \bibnamefont{Moore}},
  \bibinfo{journal}{Accounts of Chemical Research}
  \textbf{\bibinfo{volume}{34}}, \bibinfo{pages}{40} (\bibinfo{year}{2001}).

\bibitem[{\citenamefont{Cremer et~al.}(2002)\citenamefont{Cremer, Sledge, and
  Heinze}}]{cremer2002chemical}
\bibinfo{author}{\bibfnamefont{S.}~\bibnamefont{Cremer}},
  \bibinfo{author}{\bibfnamefont{M.~F.} \bibnamefont{Sledge}},
  \bibnamefont{and} \bibinfo{author}{\bibfnamefont{J.}~\bibnamefont{Heinze}},
  \bibinfo{journal}{Nature} \textbf{\bibinfo{volume}{419}},
  \bibinfo{pages}{897} (\bibinfo{year}{2002}).

\bibitem[{\citenamefont{van Kasteren et~al.}(2007)\citenamefont{van Kasteren,
  Kramer, Jensen, Campbell, Kirkpatrick, Oldham, Anthony, and
  Davis}}]{van2007expanding}
\bibinfo{author}{\bibfnamefont{S.~I.} \bibnamefont{van Kasteren}},
  \bibinfo{author}{\bibfnamefont{H.~B.} \bibnamefont{Kramer}},
  \bibinfo{author}{\bibfnamefont{H.~H.} \bibnamefont{Jensen}},
  \bibinfo{author}{\bibfnamefont{S.~J.} \bibnamefont{Campbell}},
  \bibinfo{author}{\bibfnamefont{J.}~\bibnamefont{Kirkpatrick}},
  \bibinfo{author}{\bibfnamefont{N.~J.} \bibnamefont{Oldham}},
  \bibinfo{author}{\bibfnamefont{D.~C.} \bibnamefont{Anthony}},
  \bibnamefont{and} \bibinfo{author}{\bibfnamefont{B.~G.} \bibnamefont{Davis}},
  \bibinfo{journal}{Nature} \textbf{\bibinfo{volume}{446}},
  \bibinfo{pages}{1105} (\bibinfo{year}{2007}).

\bibitem[{\citenamefont{Leininger et~al.}(2000)\citenamefont{Leininger,
  Olenyuk, and Stang}}]{leininger2000self}
\bibinfo{author}{\bibfnamefont{S.}~\bibnamefont{Leininger}},
  \bibinfo{author}{\bibfnamefont{B.}~\bibnamefont{Olenyuk}}, \bibnamefont{and}
  \bibinfo{author}{\bibfnamefont{P.~J.} \bibnamefont{Stang}},
  \bibinfo{journal}{Chemical Reviews} \textbf{\bibinfo{volume}{100}},
  \bibinfo{pages}{853} (\bibinfo{year}{2000}).

\bibitem[{\citenamefont{Zhang et~al.}(2005)\citenamefont{Zhang, Keys, Chen, and
  Glotzer}}]{zhang2005self}
\bibinfo{author}{\bibfnamefont{Z.}~\bibnamefont{Zhang}},
  \bibinfo{author}{\bibfnamefont{A.~S.} \bibnamefont{Keys}},
  \bibinfo{author}{\bibfnamefont{T.}~\bibnamefont{Chen}}, \bibnamefont{and}
  \bibinfo{author}{\bibfnamefont{S.~C.} \bibnamefont{Glotzer}},
  \bibinfo{journal}{Langmuir} \textbf{\bibinfo{volume}{21}},
  \bibinfo{pages}{11547} (\bibinfo{year}{2005}).

\bibitem[{\citenamefont{Mitov et~al.}(2002)\citenamefont{Mitov, Portet,
  Bourgerette, Snoeck, and Verelst}}]{mitov2002long}
\bibinfo{author}{\bibfnamefont{M.}~\bibnamefont{Mitov}},
  \bibinfo{author}{\bibfnamefont{C.}~\bibnamefont{Portet}},
  \bibinfo{author}{\bibfnamefont{C.}~\bibnamefont{Bourgerette}},
  \bibinfo{author}{\bibfnamefont{E.}~\bibnamefont{Snoeck}}, \bibnamefont{and}
  \bibinfo{author}{\bibfnamefont{M.}~\bibnamefont{Verelst}},
  \bibinfo{journal}{Nature materials} \textbf{\bibinfo{volume}{1}},
  \bibinfo{pages}{229} (\bibinfo{year}{2002}).

\bibitem[{\citenamefont{Whitesell et~al.}(1994)\citenamefont{Whitesell, Davis,
  Wong, and Chang}}]{whitesell1994molecular}
\bibinfo{author}{\bibfnamefont{J.~K.} \bibnamefont{Whitesell}},
  \bibinfo{author}{\bibfnamefont{R.~E.} \bibnamefont{Davis}},
  \bibinfo{author}{\bibfnamefont{M.~S.} \bibnamefont{Wong}}, \bibnamefont{and}
  \bibinfo{author}{\bibfnamefont{N.~L.} \bibnamefont{Chang}},
  \bibinfo{journal}{Journal of the American Chemical Society}
  \textbf{\bibinfo{volume}{116}}, \bibinfo{pages}{523} (\bibinfo{year}{1994}).

\bibitem[{\citenamefont{Chirik and
  Wieghardt}(2010)}]{ChirikScience327-794(2010)}
\bibinfo{author}{\bibfnamefont{P.}~\bibnamefont{Chirik}} \bibnamefont{and}
  \bibinfo{author}{\bibfnamefont{K.}~\bibnamefont{Wieghardt}},
  \bibinfo{journal}{Science} \textbf{\bibinfo{volume}{327}},
  \bibinfo{pages}{794} (\bibinfo{year}{2010}).

\bibitem[{\citenamefont{Gust et~al.}(1993)\citenamefont{Gust, Moore, and
  Moore}}]{gust1993molecular}
\bibinfo{author}{\bibfnamefont{D.}~\bibnamefont{Gust}},
  \bibinfo{author}{\bibfnamefont{T.~A.} \bibnamefont{Moore}}, \bibnamefont{and}
  \bibinfo{author}{\bibfnamefont{A.~L.} \bibnamefont{Moore}},
  \bibinfo{journal}{Accounts of Chemical Research}
  \textbf{\bibinfo{volume}{26}}, \bibinfo{pages}{198} (\bibinfo{year}{1993}).

\bibitem[{\citenamefont{De~Silva et~al.}(1997)\citenamefont{De~Silva,
  Gunaratne, Gunnlaugsson, Huxley, McCoy, Rademacher, and
  Rice}}]{de1997signaling}
\bibinfo{author}{\bibfnamefont{A.~P.} \bibnamefont{De~Silva}},
  \bibinfo{author}{\bibfnamefont{H.~N.} \bibnamefont{Gunaratne}},
  \bibinfo{author}{\bibfnamefont{T.}~\bibnamefont{Gunnlaugsson}},
  \bibinfo{author}{\bibfnamefont{A.~J.} \bibnamefont{Huxley}},
  \bibinfo{author}{\bibfnamefont{C.~P.} \bibnamefont{McCoy}},
  \bibinfo{author}{\bibfnamefont{J.~T.} \bibnamefont{Rademacher}},
  \bibnamefont{and} \bibinfo{author}{\bibfnamefont{T.~E.} \bibnamefont{Rice}},
  \bibinfo{journal}{Chemical Reviews} \textbf{\bibinfo{volume}{97}},
  \bibinfo{pages}{1515} (\bibinfo{year}{1997}).

\bibitem[{\citenamefont{Della~Gaspera et~al.}(2014)\citenamefont{Della~Gaspera,
  van Embden, Chesman, Duffy, and Jasieniak}}]{della2014mimicry}
\bibinfo{author}{\bibfnamefont{E.}~\bibnamefont{Della~Gaspera}},
  \bibinfo{author}{\bibfnamefont{J.}~\bibnamefont{van Embden}},
  \bibinfo{author}{\bibfnamefont{A.~S.} \bibnamefont{Chesman}},
  \bibinfo{author}{\bibfnamefont{N.~W.} \bibnamefont{Duffy}}, \bibnamefont{and}
  \bibinfo{author}{\bibfnamefont{J.~J.} \bibnamefont{Jasieniak}},
  \bibinfo{journal}{ACS applied materials \& interfaces}
  \textbf{\bibinfo{volume}{6}}, \bibinfo{pages}{22519} (\bibinfo{year}{2014}).

\bibitem[{\citenamefont{Rabong et~al.}(2014)\citenamefont{Rabong, Schuster,
  Liptaj, Pr{\'o}nayov{\'a}, Delchev, Jordis, and Phopase}}]{rabong2014nxo}
\bibinfo{author}{\bibfnamefont{C.}~\bibnamefont{Rabong}},
  \bibinfo{author}{\bibfnamefont{C.}~\bibnamefont{Schuster}},
  \bibinfo{author}{\bibfnamefont{T.}~\bibnamefont{Liptaj}},
  \bibinfo{author}{\bibfnamefont{N.}~\bibnamefont{Pr{\'o}nayov{\'a}}},
  \bibinfo{author}{\bibfnamefont{V.~B.} \bibnamefont{Delchev}},
  \bibinfo{author}{\bibfnamefont{U.}~\bibnamefont{Jordis}}, \bibnamefont{and}
  \bibinfo{author}{\bibfnamefont{J.}~\bibnamefont{Phopase}},
  \bibinfo{journal}{RSC Advances} \textbf{\bibinfo{volume}{4}},
  \bibinfo{pages}{21351} (\bibinfo{year}{2014}).

\bibitem[{Sup()}]{SupplementalM}
\bibinfo{note}{See Supplemental Material at ***************************}.

\bibitem[{\citenamefont{Bandrauk and Shen}(1992)}]{bandrauk1992higher}
\bibinfo{author}{\bibfnamefont{A.~D.} \bibnamefont{Bandrauk}} \bibnamefont{and}
  \bibinfo{author}{\bibfnamefont{H.}~\bibnamefont{Shen}},
  \bibinfo{journal}{Canadian Journal of Chemistry}
  \textbf{\bibinfo{volume}{70}}, \bibinfo{pages}{555} (\bibinfo{year}{1992}).

\bibitem[{\citenamefont{Gross et~al.}(1993)\citenamefont{Gross, Singh, Rabitz,
  Mease, and Huang}}]{gross1993inverse}
\bibinfo{author}{\bibfnamefont{P.}~\bibnamefont{Gross}},
  \bibinfo{author}{\bibfnamefont{H.}~\bibnamefont{Singh}},
  \bibinfo{author}{\bibfnamefont{H.}~\bibnamefont{Rabitz}},
  \bibinfo{author}{\bibfnamefont{K.}~\bibnamefont{Mease}}, \bibnamefont{and}
  \bibinfo{author}{\bibfnamefont{G.}~\bibnamefont{Huang}},
  \bibinfo{journal}{Physical Review A} \textbf{\bibinfo{volume}{47}},
  \bibinfo{pages}{4593} (\bibinfo{year}{1993}).

\bibitem[{\citenamefont{Caldeira and
  Leggett}(1983{\natexlab{a}})}]{caldeira1983path}
\bibinfo{author}{\bibfnamefont{A.~O.} \bibnamefont{Caldeira}} \bibnamefont{and}
  \bibinfo{author}{\bibfnamefont{A.~J.} \bibnamefont{Leggett}},
  \bibinfo{journal}{Physica A: Statistical mechanics and its Applications}
  \textbf{\bibinfo{volume}{121}}, \bibinfo{pages}{587}
  (\bibinfo{year}{1983}{\natexlab{a}}).

\bibitem[{\citenamefont{Caldeira and
  Leggett}(1983{\natexlab{b}})}]{caldeira1983quantum}
\bibinfo{author}{\bibfnamefont{A.}~\bibnamefont{Caldeira}} \bibnamefont{and}
  \bibinfo{author}{\bibfnamefont{A.~J.} \bibnamefont{Leggett}},
  \bibinfo{journal}{Annals of Physics} \textbf{\bibinfo{volume}{149}},
  \bibinfo{pages}{374} (\bibinfo{year}{1983}{\natexlab{b}}).

\bibitem[{\citenamefont{Gardiner et~al.}(1985)}]{gardiner1985handbook}
\bibinfo{author}{\bibfnamefont{C.~W.} \bibnamefont{Gardiner}}
  \bibnamefont{et~al.}, \emph{\bibinfo{title}{Handbook of stochastic methods}},
  vol.~\bibinfo{volume}{4} (\bibinfo{publisher}{Springer Berlin},
  \bibinfo{year}{1985}).

\bibitem[{\citenamefont{Corkum}(1993)}]{PhysRevLett.71.1994}
\bibinfo{author}{\bibfnamefont{P.~B.} \bibnamefont{Corkum}},
  \bibinfo{journal}{Phys. Rev. Lett.} \textbf{\bibinfo{volume}{71}}
  (\bibinfo{year}{1993}).

\bibitem[{\citenamefont{Winterfeldt et~al.}(2008)\citenamefont{Winterfeldt,
  Spielmann, and Gerber}}]{RevModPhys.80.117}
\bibinfo{author}{\bibfnamefont{C.}~\bibnamefont{Winterfeldt}},
  \bibinfo{author}{\bibfnamefont{C.}~\bibnamefont{Spielmann}},
  \bibnamefont{and} \bibinfo{author}{\bibfnamefont{G.}~\bibnamefont{Gerber}},
  \bibinfo{journal}{Rev. Mod. Phys.} \textbf{\bibinfo{volume}{80}}
  (\bibinfo{year}{2008}).

\bibitem[{\citenamefont{Bauer}(1997)}]{PhysRevA.56.3028}
\bibinfo{author}{\bibfnamefont{D.}~\bibnamefont{Bauer}},
  \bibinfo{journal}{Phys. Rev. A} \textbf{\bibinfo{volume}{56}}
  (\bibinfo{year}{1997}).

\bibitem[{\citenamefont{Boyd}(2008)}]{Boyd}
\bibinfo{author}{\bibfnamefont{R.~W.} \bibnamefont{Boyd}},
  \emph{\bibinfo{title}{{Nonlinear Optics, Third Edition}}}
  (\bibinfo{publisher}{Academic Press}, \bibinfo{year}{2008}).

\bibitem[{\citenamefont{Viola and Lloyd}(1998)}]{PhysRevA.58.2733}
\bibinfo{author}{\bibfnamefont{L.}~\bibnamefont{Viola}} \bibnamefont{and}
  \bibinfo{author}{\bibfnamefont{S.}~\bibnamefont{Lloyd}},
  \bibinfo{journal}{Phys. Rev. A} \textbf{\bibinfo{volume}{58}}
  (\bibinfo{year}{1998}).

\bibitem[{\citenamefont{Facchi et~al.}(2005)\citenamefont{Facchi, Tasaki,
  Pascazio, Nakazato, Tokuse, and Lidar}}]{PhysRevA.71.022302}
\bibinfo{author}{\bibfnamefont{P.}~\bibnamefont{Facchi}},
  \bibinfo{author}{\bibfnamefont{S.}~\bibnamefont{Tasaki}},
  \bibinfo{author}{\bibfnamefont{S.}~\bibnamefont{Pascazio}},
  \bibinfo{author}{\bibfnamefont{H.}~\bibnamefont{Nakazato}},
  \bibinfo{author}{\bibfnamefont{A.}~\bibnamefont{Tokuse}}, \bibnamefont{and}
  \bibinfo{author}{\bibfnamefont{D.~A.} \bibnamefont{Lidar}},
  \bibinfo{journal}{Phys. Rev. A} \textbf{\bibinfo{volume}{71}}
  (\bibinfo{year}{2005}).

\bibitem[{\citenamefont{Ban}(1998)}]{ban1998photon}
\bibinfo{author}{\bibfnamefont{M.}~\bibnamefont{Ban}},
  \bibinfo{journal}{journal of modern optics} \textbf{\bibinfo{volume}{45}},
  \bibinfo{pages}{2315} (\bibinfo{year}{1998}).

\bibitem[{\citenamefont{Cappellaro et~al.}(2006)\citenamefont{Cappellaro,
  Hodges, Havel, and Cory}}]{cappellaro2006principles}
\bibinfo{author}{\bibfnamefont{P.}~\bibnamefont{Cappellaro}},
  \bibinfo{author}{\bibfnamefont{J.}~\bibnamefont{Hodges}},
  \bibinfo{author}{\bibfnamefont{T.}~\bibnamefont{Havel}}, \bibnamefont{and}
  \bibinfo{author}{\bibfnamefont{D.}~\bibnamefont{Cory}}, \bibinfo{journal}{The
  Journal of chemical physics} \textbf{\bibinfo{volume}{125}},
  \bibinfo{pages}{044514} (\bibinfo{year}{2006}).

\bibitem[{\citenamefont{Constant et~al.}(1999)\citenamefont{Constant, Garzella,
  Breger, M\'evel, Dorrer, Le~Blanc, Salin, and
  Agostini}}]{Constant1999PhysRevLett.82.1668}
\bibinfo{author}{\bibfnamefont{E.}~\bibnamefont{Constant}},
  \bibinfo{author}{\bibfnamefont{D.}~\bibnamefont{Garzella}},
  \bibinfo{author}{\bibfnamefont{P.}~\bibnamefont{Breger}},
  \bibinfo{author}{\bibfnamefont{E.}~\bibnamefont{M\'evel}},
  \bibinfo{author}{\bibfnamefont{C.}~\bibnamefont{Dorrer}},
  \bibinfo{author}{\bibfnamefont{C.}~\bibnamefont{Le~Blanc}},
  \bibinfo{author}{\bibfnamefont{F.}~\bibnamefont{Salin}}, \bibnamefont{and}
  \bibinfo{author}{\bibfnamefont{P.}~\bibnamefont{Agostini}},
  \bibinfo{journal}{Phys. Rev. Lett.} \textbf{\bibinfo{volume}{82}},
  \bibinfo{pages}{1668} (\bibinfo{year}{1999}).

\bibitem[{\citenamefont{Franco et~al.}(2010)\citenamefont{Franco, Spanner, and
  Brumer}}]{Franco2010143}
\bibinfo{author}{\bibfnamefont{I.}~\bibnamefont{Franco}},
  \bibinfo{author}{\bibfnamefont{M.}~\bibnamefont{Spanner}}, \bibnamefont{and}
  \bibinfo{author}{\bibfnamefont{P.}~\bibnamefont{Brumer}},
  \bibinfo{journal}{Chemical Physics} \textbf{\bibinfo{volume}{370}},
  \bibinfo{pages}{143 } (\bibinfo{year}{2010}).

\bibitem[{\citenamefont{Schwieters and Rabitz}(1993)}]{schwieters1993optimal}
\bibinfo{author}{\bibfnamefont{C.~D.} \bibnamefont{Schwieters}}
  \bibnamefont{and} \bibinfo{author}{\bibfnamefont{H.}~\bibnamefont{Rabitz}},
  \bibinfo{journal}{Physical Review A} \textbf{\bibinfo{volume}{48}},
  \bibinfo{pages}{2549} (\bibinfo{year}{1993}).

\bibitem[{\citenamefont{Blackburn et~al.}(2016)\citenamefont{Blackburn,
  Cirillo, and Gr{\o}nbech-Jensen}}]{blackburn2016survey}
\bibinfo{author}{\bibfnamefont{J.~A.} \bibnamefont{Blackburn}},
  \bibinfo{author}{\bibfnamefont{M.}~\bibnamefont{Cirillo}}, \bibnamefont{and}
  \bibinfo{author}{\bibfnamefont{N.}~\bibnamefont{Gr{\o}nbech-Jensen}},
  \bibinfo{journal}{Physics Reports} \textbf{\bibinfo{volume}{611}},
  \bibinfo{pages}{1} (\bibinfo{year}{2016}).

\bibitem[{\citenamefont{Cabrera et~al.}(2015)\citenamefont{Cabrera, Bondar,
  Jacobs, and Rabitz}}]{PhysRevA.92.042122}
\bibinfo{author}{\bibfnamefont{R.}~\bibnamefont{Cabrera}},
  \bibinfo{author}{\bibfnamefont{D.~I.} \bibnamefont{Bondar}},
  \bibinfo{author}{\bibfnamefont{K.}~\bibnamefont{Jacobs}}, \bibnamefont{and}
  \bibinfo{author}{\bibfnamefont{H.~A.} \bibnamefont{Rabitz}},
  \bibinfo{journal}{Phys. Rev. A} \textbf{\bibinfo{volume}{92}},
  \bibinfo{pages}{042122} (\bibinfo{year}{2015}).

\bibitem[{\citenamefont{Andrews and Bradshaw}(2004)}]{andrews2004virtual}
\bibinfo{author}{\bibfnamefont{D.~L.} \bibnamefont{Andrews}} \bibnamefont{and}
  \bibinfo{author}{\bibfnamefont{D.~S.} \bibnamefont{Bradshaw}},
  \bibinfo{journal}{European journal of physics} \textbf{\bibinfo{volume}{25}},
  \bibinfo{pages}{845} (\bibinfo{year}{2004}).

\bibitem[{\citenamefont{Duque et~al.}(2015)\citenamefont{Duque, Brumer, and
  Pach\'on}}]{PhysRevLett.115.110402Brumer}
\bibinfo{author}{\bibfnamefont{S.}~\bibnamefont{Duque}},
  \bibinfo{author}{\bibfnamefont{P.}~\bibnamefont{Brumer}}, \bibnamefont{and}
  \bibinfo{author}{\bibfnamefont{L.~A.} \bibnamefont{Pach\'on}},
  \bibinfo{journal}{Phys. Rev. Lett.} \textbf{\bibinfo{volume}{115}},
  \bibinfo{pages}{110402} (\bibinfo{year}{2015}).

\bibitem[{\citenamefont{Li et~al.}(2002)\citenamefont{Li, Turinici,
  Ramakrishna, and Rabitz}}]{li2002optimal}
\bibinfo{author}{\bibfnamefont{B.}~\bibnamefont{Li}},
  \bibinfo{author}{\bibfnamefont{G.}~\bibnamefont{Turinici}},
  \bibinfo{author}{\bibfnamefont{V.}~\bibnamefont{Ramakrishna}},
  \bibnamefont{and} \bibinfo{author}{\bibfnamefont{H.}~\bibnamefont{Rabitz}},
  \bibinfo{journal}{The Journal of Physical Chemistry B}
  \textbf{\bibinfo{volume}{106}}, \bibinfo{pages}{8125} (\bibinfo{year}{2002}).

\bibitem[{\citenamefont{Goun et~al.}(2016)\citenamefont{Goun, Bondar, Ali,
  Quine, and Rabitz}}]{goun2016photonic}
\bibinfo{author}{\bibfnamefont{A.}~\bibnamefont{Goun}},
  \bibinfo{author}{\bibfnamefont{D.~I.} \bibnamefont{Bondar}},
  \bibinfo{author}{\bibfnamefont{O.~E.} \bibnamefont{Ali}},
  \bibinfo{author}{\bibfnamefont{Z.}~\bibnamefont{Quine}}, \bibnamefont{and}
  \bibinfo{author}{\bibfnamefont{H.~A.} \bibnamefont{Rabitz}},
  \bibinfo{journal}{Scientific reports} \textbf{\bibinfo{volume}{6}},
  \bibinfo{pages}{25827} (\bibinfo{year}{2016}).

\bibitem[{\citenamefont{Brixner et~al.}(2001)\citenamefont{Brixner, Damrauer,
  Niklaus, and Gerber}}]{brixner2001photoselective}
\bibinfo{author}{\bibfnamefont{T.}~\bibnamefont{Brixner}},
  \bibinfo{author}{\bibfnamefont{N.}~\bibnamefont{Damrauer}},
  \bibinfo{author}{\bibfnamefont{P.}~\bibnamefont{Niklaus}}, \bibnamefont{and}
  \bibinfo{author}{\bibfnamefont{G.}~\bibnamefont{Gerber}},
  \bibinfo{journal}{Nature} \textbf{\bibinfo{volume}{414}}, \bibinfo{pages}{57}
  (\bibinfo{year}{2001}).

\bibitem[{\citenamefont{Tkaczyk et~al.}(2008)\citenamefont{Tkaczyk, Mauring,
  Tkaczyk, Krasnenko, Ye, Baker, and Norris}}]{tkaczyk2008control}
\bibinfo{author}{\bibfnamefont{E.~R.} \bibnamefont{Tkaczyk}},
  \bibinfo{author}{\bibfnamefont{K.}~\bibnamefont{Mauring}},
  \bibinfo{author}{\bibfnamefont{A.~H.} \bibnamefont{Tkaczyk}},
  \bibinfo{author}{\bibfnamefont{V.}~\bibnamefont{Krasnenko}},
  \bibinfo{author}{\bibfnamefont{J.~Y.} \bibnamefont{Ye}},
  \bibinfo{author}{\bibfnamefont{J.~R.} \bibnamefont{Baker}}, \bibnamefont{and}
  \bibinfo{author}{\bibfnamefont{T.~B.} \bibnamefont{Norris}},
  \bibinfo{journal}{Biochemical and biophysical research communications}
  \textbf{\bibinfo{volume}{376}}, \bibinfo{pages}{733} (\bibinfo{year}{2008}).

\bibitem[{\citenamefont{Roth et~al.}(2009)\citenamefont{Roth, Guyon, Roslund,
  Boutou, Courvoisier, Wolf, and Rabitz}}]{roth2009quantum}
\bibinfo{author}{\bibfnamefont{M.}~\bibnamefont{Roth}},
  \bibinfo{author}{\bibfnamefont{L.}~\bibnamefont{Guyon}},
  \bibinfo{author}{\bibfnamefont{J.}~\bibnamefont{Roslund}},
  \bibinfo{author}{\bibfnamefont{V.}~\bibnamefont{Boutou}},
  \bibinfo{author}{\bibfnamefont{F.}~\bibnamefont{Courvoisier}},
  \bibinfo{author}{\bibfnamefont{J.-P.} \bibnamefont{Wolf}}, \bibnamefont{and}
  \bibinfo{author}{\bibfnamefont{H.}~\bibnamefont{Rabitz}},
  \bibinfo{journal}{Physical review letters} \textbf{\bibinfo{volume}{102}},
  \bibinfo{pages}{253001} (\bibinfo{year}{2009}).

\bibitem[{\citenamefont{Petersen et~al.}(2010)\citenamefont{Petersen,
  Mitri{\'c}, Bona{\v{c}}i{\'c}-Kouteck{\`y}, Wolf, Roslund, and
  Rabitz}}]{petersen2010shaped}
\bibinfo{author}{\bibfnamefont{J.}~\bibnamefont{Petersen}},
  \bibinfo{author}{\bibfnamefont{R.}~\bibnamefont{Mitri{\'c}}},
  \bibinfo{author}{\bibfnamefont{V.}~\bibnamefont{Bona{\v{c}}i{\'c}-Kouteck{\`y}}},
  \bibinfo{author}{\bibfnamefont{J.-P.} \bibnamefont{Wolf}},
  \bibinfo{author}{\bibfnamefont{J.}~\bibnamefont{Roslund}}, \bibnamefont{and}
  \bibinfo{author}{\bibfnamefont{H.}~\bibnamefont{Rabitz}},
  \bibinfo{journal}{Physical review letters} \textbf{\bibinfo{volume}{105}},
  \bibinfo{pages}{073003} (\bibinfo{year}{2010}).

\bibitem[{\citenamefont{Rondi et~al.}(2011)\citenamefont{Rondi, Kiselev,
  Machado, Extermann, Weber, Bonacina, Wolf, Roslund, Roth, and
  Rabitz}}]{rondi2011discriminating}
\bibinfo{author}{\bibfnamefont{A.}~\bibnamefont{Rondi}},
  \bibinfo{author}{\bibfnamefont{D.}~\bibnamefont{Kiselev}},
  \bibinfo{author}{\bibfnamefont{S.}~\bibnamefont{Machado}},
  \bibinfo{author}{\bibfnamefont{J.}~\bibnamefont{Extermann}},
  \bibinfo{author}{\bibfnamefont{S.}~\bibnamefont{Weber}},
  \bibinfo{author}{\bibfnamefont{L.}~\bibnamefont{Bonacina}},
  \bibinfo{author}{\bibfnamefont{J.-P.} \bibnamefont{Wolf}},
  \bibinfo{author}{\bibfnamefont{J.}~\bibnamefont{Roslund}},
  \bibinfo{author}{\bibfnamefont{M.}~\bibnamefont{Roth}}, \bibnamefont{and}
  \bibinfo{author}{\bibfnamefont{H.}~\bibnamefont{Rabitz}},
  \bibinfo{journal}{CHIMIA International Journal for Chemistry}
  \textbf{\bibinfo{volume}{65}}, \bibinfo{pages}{346} (\bibinfo{year}{2011}).

\bibitem[{\citenamefont{Rondi et~al.}(2012)\citenamefont{Rondi, Bonacina,
  Trisorio, Hauri, and Wolf}}]{rondi2012coherent}
\bibinfo{author}{\bibfnamefont{A.}~\bibnamefont{Rondi}},
  \bibinfo{author}{\bibfnamefont{L.}~\bibnamefont{Bonacina}},
  \bibinfo{author}{\bibfnamefont{A.}~\bibnamefont{Trisorio}},
  \bibinfo{author}{\bibfnamefont{C.}~\bibnamefont{Hauri}}, \bibnamefont{and}
  \bibinfo{author}{\bibfnamefont{J.-P.} \bibnamefont{Wolf}},
  \bibinfo{journal}{Physical Chemistry Chemical Physics}
  \textbf{\bibinfo{volume}{14}}, \bibinfo{pages}{9317} (\bibinfo{year}{2012}).

\bibitem[{\citenamefont{Feng et~al.}(2000)\citenamefont{Feng, Mellor,
  Bernstein, Keller-Peck, Nguyen, Wallace, Nerbonne, Lichtman, and
  Sanes}}]{feng2000imaging}
\bibinfo{author}{\bibfnamefont{G.}~\bibnamefont{Feng}},
  \bibinfo{author}{\bibfnamefont{R.~H.} \bibnamefont{Mellor}},
  \bibinfo{author}{\bibfnamefont{M.}~\bibnamefont{Bernstein}},
  \bibinfo{author}{\bibfnamefont{C.}~\bibnamefont{Keller-Peck}},
  \bibinfo{author}{\bibfnamefont{Q.~T.} \bibnamefont{Nguyen}},
  \bibinfo{author}{\bibfnamefont{M.}~\bibnamefont{Wallace}},
  \bibinfo{author}{\bibfnamefont{J.~M.} \bibnamefont{Nerbonne}},
  \bibinfo{author}{\bibfnamefont{J.~W.} \bibnamefont{Lichtman}},
  \bibnamefont{and} \bibinfo{author}{\bibfnamefont{J.~R.} \bibnamefont{Sanes}},
  \bibinfo{journal}{Neuron} \textbf{\bibinfo{volume}{28}}, \bibinfo{pages}{41}
  (\bibinfo{year}{2000}).

\bibitem[{\citenamefont{Bagwell and Adams}(1993)}]{bagwell1993fluorescence}
\bibinfo{author}{\bibfnamefont{C.~B.} \bibnamefont{Bagwell}} \bibnamefont{and}
  \bibinfo{author}{\bibfnamefont{E.~G.} \bibnamefont{Adams}},
  \bibinfo{journal}{Annals of the New York Academy of Sciences}
  \textbf{\bibinfo{volume}{677}}, \bibinfo{pages}{167} (\bibinfo{year}{1993}).

\bibitem[{\citenamefont{Speicher et~al.}(1996)\citenamefont{Speicher, Ballard,
  and Ward}}]{speicher1996karyotyping}
\bibinfo{author}{\bibfnamefont{M.~R.} \bibnamefont{Speicher}},
  \bibinfo{author}{\bibfnamefont{S.~G.} \bibnamefont{Ballard}},
  \bibnamefont{and} \bibinfo{author}{\bibfnamefont{D.~C.} \bibnamefont{Ward}},
  \bibinfo{journal}{Nature genetics} \textbf{\bibinfo{volume}{12}},
  \bibinfo{pages}{368} (\bibinfo{year}{1996}).

\bibitem[{\citenamefont{Perfetto et~al.}(2004)\citenamefont{Perfetto,
  Chattopadhyay, and Roederer}}]{perfetto2004seventeen}
\bibinfo{author}{\bibfnamefont{S.~P.} \bibnamefont{Perfetto}},
  \bibinfo{author}{\bibfnamefont{P.~K.} \bibnamefont{Chattopadhyay}},
  \bibnamefont{and} \bibinfo{author}{\bibfnamefont{M.}~\bibnamefont{Roederer}},
  \bibinfo{journal}{Nature Reviews Immunology} \textbf{\bibinfo{volume}{4}},
  \bibinfo{pages}{648} (\bibinfo{year}{2004}).

\bibitem[{\citenamefont{Sommer et~al.}(2016)\citenamefont{Sommer, Bothschafter,
  Sato, Jakubeit, Latka, Razskazovskaya, Fattahi, Jobst, Schweinberger,
  Shirvanyan et~al.}}]{sommer2016attosecond}
\bibinfo{author}{\bibfnamefont{A.}~\bibnamefont{Sommer}},
  \bibinfo{author}{\bibfnamefont{E.}~\bibnamefont{Bothschafter}},
  \bibinfo{author}{\bibfnamefont{S.}~\bibnamefont{Sato}},
  \bibinfo{author}{\bibfnamefont{C.}~\bibnamefont{Jakubeit}},
  \bibinfo{author}{\bibfnamefont{T.}~\bibnamefont{Latka}},
  \bibinfo{author}{\bibfnamefont{O.}~\bibnamefont{Razskazovskaya}},
  \bibinfo{author}{\bibfnamefont{H.}~\bibnamefont{Fattahi}},
  \bibinfo{author}{\bibfnamefont{M.}~\bibnamefont{Jobst}},
  \bibinfo{author}{\bibfnamefont{W.}~\bibnamefont{Schweinberger}},
  \bibinfo{author}{\bibfnamefont{V.}~\bibnamefont{Shirvanyan}},
  \bibnamefont{et~al.}, \bibinfo{journal}{Nature}
  \textbf{\bibinfo{volume}{534}}, \bibinfo{pages}{86} (\bibinfo{year}{2016}).

\end{thebibliography}

\end{document}